# Broadband Quantum Efficiency Enhancement in High Index Nanowires Resonators

*Yiming Yang,[1] Xingyue Peng,[1] Steven Hyatt,[1] and Dong Yu[1*]*

[1]Department of Physics, University of California, Davis, CA 95616, USA

Light trapping in sub-wavelength semiconductor nanowires (NWs) offers a promising approach to simultaneously reducing material consumption and enhancing photovoltaic performance[1-7]. Nevertheless, the absorption efficiency of a NW, defined by the ratio of optical absorption cross section to the NW diameter, lingers around 1 in existing NW photonic devices[1,5-10], and the absorption enhancement suffers from a narrow spectral width. Here, we show that the absorption efficiency can be significantly improved in NWs with higher refractive indices, by an experimental observation of up to 350% external quantum efficiency (EQE) in lead sulfide (PbS) NW resonators, a 3-fold increase compared to Si NWs[5-7,10]. Furthermore, broadband absorption enhancement is achieved in single tapered NWs, where light of various wavelengths is absorbed at segments with different diameters analogous to a tandem solar cell. Overall, the single NW Schottky junction solar cells benefit from optical resonance, near bandgap open circuit voltage, and long minority carrier diffusion length, demonstrating power conversion efficiency (PCE) **comparable to single Si NW coaxial p-n junction cells[11], but with much simpler fabrication processes.**

Semiconductor NW solar cells have demonstrated promising potentials in solar energy conversion[12-16], benefiting from their low-cost fabrication, efficient charge separation at the NW interfaces, and increased optical absorption through light confinement[1,3-7,9-13,17-22]. As the NW diameter is often shorter than the light wavelength, ray optics fails to describe the optical absorption in the NWs, and the wave nature of light has to be considered. As a result, resonance absorption in sub-wavelength NWs arises, which can be understood by the formation of standing waves in the NW waveguide. At resonance frequencies, the optical absorption cross section of a NW ($\sigma_{abs}$, defined by $P_{abs}/I_0$, where $P_{abs}$ is the absorbed power per unit NW length and $I_0$ is the light intensity) may expand far beyond its diameter ($d_{NW}$). This absorption enhancement has attracted much attention with hope to increase the optical absorption of the NW arrays, which can lower material consumption and reduce charge recombination. Notably, because the spectral width of the resonance peak is often much narrower than the solar spectrum, broadband absorption enhancement through geometric design and material engineering is necessary for improving the overall solar energy conversion efficiency.

Resonance absorption is often evaluated in planar devices[1,7,9,10,23] where NWs are oriented parallel to the substrates (Figure 1a). In this configuration, the absorption efficiency, defined as $Q_{abs} = \sigma_{abs}/d_{NW}$, can be calculated by numerically solving Maxwell equations and compared to the experimentally determined EQE, obtained by only counting the incident photons projected to the physical size of the NW[1]. If $\sigma_{abs}$ exceeds $d_{NW}$ because of resonance, over-unity $Q_{abs}$ and EQE are expected. Unfortunately, the existing planar NW resonators exhibit EQEs only slightly over unity. To date, EQE of 120% has been observed in planar Si NWs[10], while theoretically calculated $Q_{abs}$ in planar Si[5-7,10], Ge[1,8,9] and GaAs[24] NW devices typically falls below 2. Furthermore, there has been no report of efficient NW resonators that have resonance peaks in the near-infrared, which contains half of the solar energy. To enhance resonance absorption, a rational approach is to employ high refractive index ($n$) materials, as the light intensity and hence the absorption in the NWs increases with $n^2$, known as the Yablonovitch limit[25]. Furthermore, optical absorption of sub-wavelength objects may surpass the thin film (Yablonovitch) limit[25,26], owing to their resonant light absorption abilities[1-7]. Here, we investigate the resonance absorption in planar devices consisting of single PbS NWs with a high refractive index both in real ($n$) and imaginary ($\kappa$) parts[27]. PbS has $n = 4.5$ and $\kappa = 1.5$ at 700 nm, in comparison to Si which has $n = 3.5$ and $\kappa < 0.1$ ($n$ and $\kappa$ values of several common materials are compared in Figure S1 in the Supplementary Information). We report a record-high EQE up to $350 \pm 30\%$ in a PbS NW with $d_{NW} = 55$ nm. PbS NWs, particularly those with larger diameters, exhibit strong resonance absorption in the near-infrared. We also demonstrate potentials for broadband enhancement in tapered NWs. Our single PbS NW Schottky solar cells exhibit a high open circuit voltage ($V_{oc}$) of 0.25 V, which can be further increased to 0.31 V by electrostatic gating, close to the bandgap of PbS (0.41 eV). Finally, taking advantage of long minority carrier diffusion lengths, our single NW Schottky solar cells show power conversion efficiency (PCE) of 3.9%, in the absence of complex core-shell NW junctions.

The PbS NWs with uniform diameters are synthesized via a vapor-liquid-solid (VLS) approach, catalyzed by Ti thin films[28]. Planar single NW devices are then fabricated with one Schottky and one Ohmic contact (see more details in *Methods*). EQEs are carefully evaluated based on the short-circuit photocurrent ($I_{sc}$) under the illumination of a linearly polarized (**E**//NW axis) laser tightly focused on the NW near the Schottky contact. The EQE is calculated by counting the photons projected into the NW from a Gaussian beam. The detailed calculation of EQE and the error analysis can be found in the Supplementary Information (SI). The wavelength-dependent EQE measured in a PbS NW of $d_{NW}$ = 55 nm is plotted in Figure 1c, with a peak value of 350 ± 30% at around 625 nm. This EQE value indicates that the optical absorption cross section is strongly enhanced to over three times larger than $d_{NW}$. We also show EQE values measured in another NW with $d_{NW}$ = 90 nm in Figure 1d, where EQE exceeds 220% at $\lambda$ = 1600 nm, demonstrating strong resonance absorption in the near infrared. The peak in the visible spectrum clearly red shifts as the diameter increases from 60 nm to 90 nm. Upon the measured more than eight devices with diameters ranging from 55 to 100 nm, the peak EQE values range from 150% to 350% varying from device to device, with five devices showing EQE values larger than 200%. We note here that multiple exciton generation (MEG) efficiency in PbS is expected to be below 10% in our measured wavelength range[29] and hence its effect can be ignored. Our measured EQEs in PbS NWs are significantly increased compared to planar Si[5-7,21] and Ge[1,16,23] NWs and for the first time demonstrate the absorption enhancement in the near infrared.

To understand the observed high EQEs, finite-difference time-domain (FDTD) simulations are performed to calculate $Q_{abs}$ as a function of $d_{NW}$ (Figure 1b). $Q_{abs}$ as high as 4.47 is obtained, which peaks at $d_{NW}$ = 35 nm and $\lambda$ = 628 nm. In NWs with larger diameters ($d_{NW}$ > 90 nm), the maximum $Q_{abs}$ exceeds 2 in the near infrared, consistent with the experimental observation (Figure 1d). As the NW diameter increases, both the visible and infrared peaks red shift, which clearly reflects the resonance nature. The simulation also predicts that PbTe NWs of higher $n$ and $\kappa$ values exhibit larger $Q_{abs}$ (Figure S4d). Our measured EQE values quantitatively agree with the calculated $Q_{abs}$ (Figures 1c,d), which indicates the photogenerated electrons and holes are nearly completely separated and collected. However, the EQE near 500 nm is significantly lower than $Q_{abs}$ in the $d_{NW}$ = 90 nm NW (Figure 1d). Similar behaviors have been observed in several other devices with large diameters. The origin is not yet clear but may be caused by faster charge recombination at shorter wavelength. At about 500 nm, the absorption is dominated by higher order resonance mode where the photogenerated carriers are closer to the surface[1] and may suffer from fast surface recombination. Such deviation is not observed in the thinner NWs because the higher order mode is out of the measured spectral range. Further investigation is necessary to fully understand this interesting behavior.

Though high EQE values are demonstrated above, the absorption peaks in a single NW with a uniform diameter have narrow bandwidths, not suitable to efficiently absorb the entire solar spectrum. Here we demonstrate a simple and practical design based on a tapered NW that substantially broadens the absorption spectrum. In this design (Figure 2a), light of different wavelength is absorbed at different segments along the NW axis, analogous to multi-junction solar cells but with much simpler fabrication. Figure 2b shows that the resonance peaks of NWs with three representative diameters can effectively cover the entire solar spectrum including the near infrared. To experimentally demonstrate broadband absorption, we have synthesized tapered PbS NWs by gradually reducing the temperature during the growth (Figure 2c). As shown in Figures 2c-d, the diameter of the NW gradually increases from 100 nm near the left contact to 220 nm in 5 μm along the NW axis. We investigate the tapered NW with scanning photocurrent microscopy (SPCM), where we raster scan a focused laser beam and measure photocurrent as a function of laser position[28]. Spatially resolved photocurrent is mapped at various laser wavelengths. At 500 nm, the photocurrent peak is near the left contact (Figures 2e-f). However, as the wavelength increases to 850 nm, the photocurrent peak position gradually shifts to the right by about 2 μm. This shift is because the larger diameter towards the right side of the NW favors the resonance absorption at longer wavelengths. Though carriers injected a distance away from the contact suffer from recombination, the increased absorption at longer wavelength dominates and leads to a stronger photocurrent. Therefore, this observation clearly demonstrates the possibilities to achieve broadband absorption by engineering the shape of the NWs.

After demonstrating the strong and broadband resonance enhancement, we now investigate the charge separation at the NW junction. The simple design of the Schottky NW devices allows a precise control of band bending by applying a back gate voltage ($V_g$). In the dark, the conductance under forward source-drain bias ($V_{sd}$) increases at negative $V_g$, indicating a p-type doping (Figure 3a). We then focus the laser at the Schottky junction and scan $V_g$ and $V_{sd}$ while measuring current (Figure 3b), from which $V_{oc}$ and $I_{sc}$ can be extracted as a function of $V_g$. We find that both $V_{oc}$ and $I_{sc}$ values can be enhanced at positive $V_g$ (Figure 3c). This can be understood by considering the field effects on the Schottky junction. A positive $V_g$ increases the depletion width and effectively reduces the leakage current at the junction. The depletion layer is equivalent to the i-layer in a p-i-n junction and $V_g$ can be used to control its width to reduce the leakage current. However, as $V_g$ is further increased above +5 V, $I_{sc}$ rapidly decreases and $V_{oc}$ also slightly decreases, caused by reduced band bending and higher NW resistance (Figure 3e). At negative $V_g$, although the Schottky band bending is increased, both $V_{oc}$ and $I_{sc}$ decrease substantially (Figure 3c), because the negative $V_g$ narrows the depletion width and increases the tunneling current (Figure 3e). The significantly increased dark saturation current under reversed bias at negative $V_g$ has been confirmed experimentally as shown in Figure S6.

The above measurements provide crucial guidelines for optimizing the cell performance through the control of junction band bending. While the junction improvement is demonstrated by a fine tuning of $V_g$ in the planar NW devices, the optimization in real NW photovoltaic devices can be achieved by adjusting the doping concentration in the NWs. Remarkably, we have observed a $V_{oc}$ = 0.25 V at $V_g$ = 0 V at an illumination intensity of 50 W/cm$^2$, which can be further increased to 0.31 V at $V_g$ = 5 V (Figure 3c), close to the PbS bandgap (0.41 eV). This high $V_{oc}$ indicates low charge recombination at the Schottky junction. $V_{oc}$ increases logarithmically with intensity as shown in Figure 3d, in good agreement with $V_{oc} = (Ak_BT/q)$ $\ln(J_{sc}/J_s +1)$, where $A$ is the ideality factor, $J_{sc}$ is the short circuit current density, and $J_s$ is the saturation current density. We note that a planar Schottky NW device acts as a solar concentrator, where the carriers injected into the NW diffuse to the Schottky junction (Figure 4b), leading to a high $J_{sc}$ and an enhanced $V_{oc}$.

Finally, we demonstrate the overall power conversion efficiency of our Schottky-junction based PbS NW solar cells. Because of the short depletion widths and minority carrier diffusion lengths ($l_D$) in most nanostructured solar materials, core-shell structures are often necessary to extend the photo-sensitive region and increase the output power[11]. Nevertheless, complex synthetic procedures and multiple-step lithographic processes are required to fabricate devices containing core-shell junctions[14,15]. The large area of the radial p-n junction also demands nontrivial interfacial engineering such as sandwiching proper insulating layers. As an alternative method, we can achieve almost complete charge collection in the absence of radial p-n junctions by taking advantage of the long $l_D$ in PbS NWs (Figure 4b). In order to achieve efficient charge collection, a $l_D$ comparable the device channel length ($l$) is required. The charge collection efficiency is $\frac{l_D}{l}\left[1 - \exp\left(-\frac{l}{l_D}\right)\right]$, which reaches 40% when $\frac{l_D}{l}$ = 0.5 and 60% when $\frac{l_D}{l}$ = 1 (see a plot of charge collection efficiency vs. $\frac{l_D}{l}$ in SI). Thanks to the long recombination lifetime in bulk PbS[28] and the well passivated NW surface[30], a $l_D$ up to 7.3 μm is confirmed by using SPCM (Figure 4a). This $l_D$ can be further increased by applying $V_g$, resulting in a nearly uniform photocurrent along the entire NW[28] (Figure 4a). This suggests that collection efficiency can be improved close to 100% by lowering the doping concentration. We estimate a PCE of 3.9% under the global illumination of the microscope light (halogen lamp at 3300 K and 0.7 W/cm$^2$). Note that the PCE is calculated using the NW diameter and channel length following other reports[10,11]. Our PCE values are comparable with other single NW solar cells with core-shell junctions,[11,23,31,32] while our device fabrication is much easier.

In summary, we have demonstrated up to 350% EQE in single PbS NW photovoltaic devices. The high efficiency originates from strong absorption enhancement and efficient charge collection at the Schottky junction. Field effects can increase the $V_{oc}$ close to bandgap by limiting the tunneling current. The long minority carrier diffusion lengths enable a simple design of Schottky NW solar cells without using more complex core-shell p-n junctions. Broadband absorption enhancement may also be realized using tapered NWs. This work provides valuable insights for optimizing light absorption and solar conversion efficiency by harnessing the resonance absorption, particularly of high refractive index materials and engineering the shape and doping concentration of semiconductor NWs. Though focused on single NW solar cells, the demonstrated concepts of Schottky devices may also be applied to ensemble solar cells based on vertical NW arrays.

**Methods**

PbS NWs were synthesized by a chemical vapor deposition (CVD) method inside a tube furnace. Details of the growth can be found elsewhere[28]. Doping of the NWs was achieved by tuning the mass ratio between the lead chloride ($PbCl_2$) and sulfur (S) precursors. The as-grown NWs were transferred to Si substrates coated with 300 nm $SiO_2$. Subsequently, electrodes were defined through electron beam lithography. The contact areas of the NWs were etched in buffered oxide etcher (BOE, 6:1) for 5 s before deposition of 75 nm Cr followed by 100nm Au. Asymmetric contacts in the one-sided Schottky devices are achieved by using controlled fast lift-off. The tension between a NW and contacts during the fast lift-off often results in a Schottky junction at one of the contacts.

Current-voltage ($I$–$V_{sd}$) characterizations were obtained through a current preamplifier (DL Instruments, model 1211) and a National Instrument (NI) data acquisition system, with gate voltage applied by a voltage amplifier (Trek). Spatially resolved optoelectronic measurements were performed on a home-built SPCM platform. A linear polarized laser with tunable wavelength (500 - 850 nm, 1100 - 1700 nm, NKT SuperK plus) was focused by a 100× N.A. 0.95 objective lens and raster scanned on a planar NW device by a pair of mirrors mounted on galvanometers. Both the reflection and the photocurrent signals were simultaneously recorded to produce a two-dimensional (2D) map of photocurrent. The excitation laser power was controlled by a set of neutral density (ND) filters and was measured by a power meter underneath the objective lens. The polarization of the laser was controlled by a half wave plate mounted on a motorized rotation stage.

FDTD simulation is performed using commercial software (Lumerical). The complex refractive index of PbS was obtained from reference 26. Plane wave sources are used in the simulation, with polarization along the axis of the NW as in the experiment. The wavelength-dependent $\sigma_{abs}$ is simulated through a wave pulse generated by the source. Convergence tests including pulse width, simulation time/region, source location/dimension, and mesh sizes were performed to ensure the validity of the simulation.

**Acknowledgements**

This work was supported by the U.S. National Science Foundation Grant DMR-1310678. Work at the Molecular Foundry was supported by the Office of Science, Office of Basic Energy Sciences, of the U.S. Department of Energy under Contract No. DE-AC02-05CH11231. We thank Nima Maghoul, Teresa Le, Xin Sun, Wenyang Li, and Jiao Li for useful discussion.

**Author Contributions**

Y. Y. and D. Y. conceived and designed the experiments and wrote the paper. X. P. and S.H. assisted in FDTD simulation and experimental measurements. All authors discussed the results and commented on the manuscript.

**Additional Information**

Supplementary information is available in the online version of the paper. Reprints and premissions information is available online at www.nature.com/reprints. Correspondence should be addressed to D.Y.

**Competing financial interests**

The authors declare no competing financial interests.

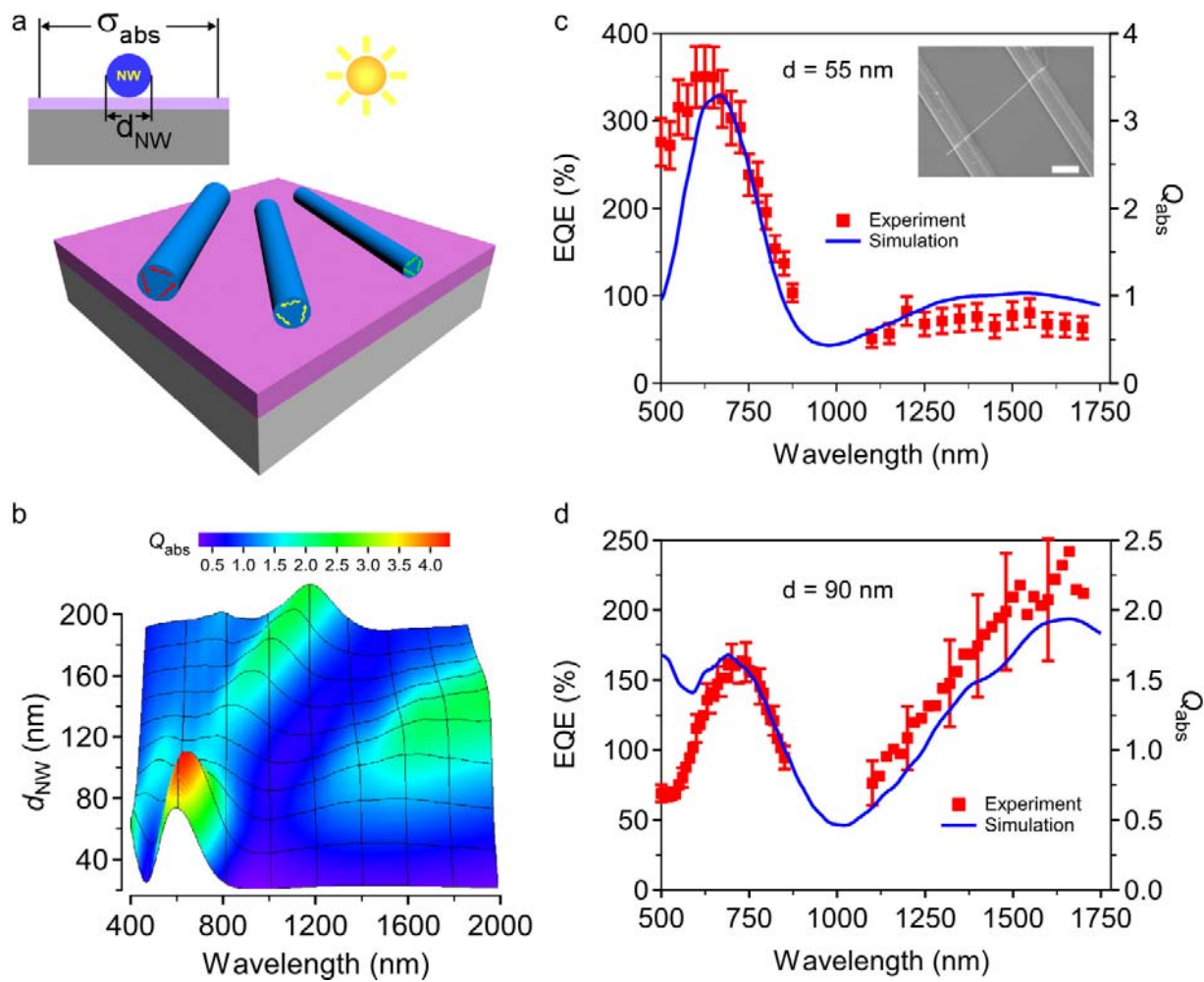

**Figure 1 | Resonance absorption in PbS NWs. a**, Schematic of resonance absorption in NWs, where the optical cross section $\sigma_{abs}$ can be enhanced beyond the NW diameter $d_{NW}$. **b**, FDTD simulation of absorption efficiency $Q_{abs}$ for PbS NWs with different diameters. **c-d,** Experimentally determined EQE (square) agrees quantitatively with the simulated $Q_{abs}$ for NWs of $d_{NW}$ = 55 and 90 nm, respectively. Scale bar in the inset is 3μm. The dashed lines indicate the resonance peaks in the visible, which clearly red shift with increasing diameters.

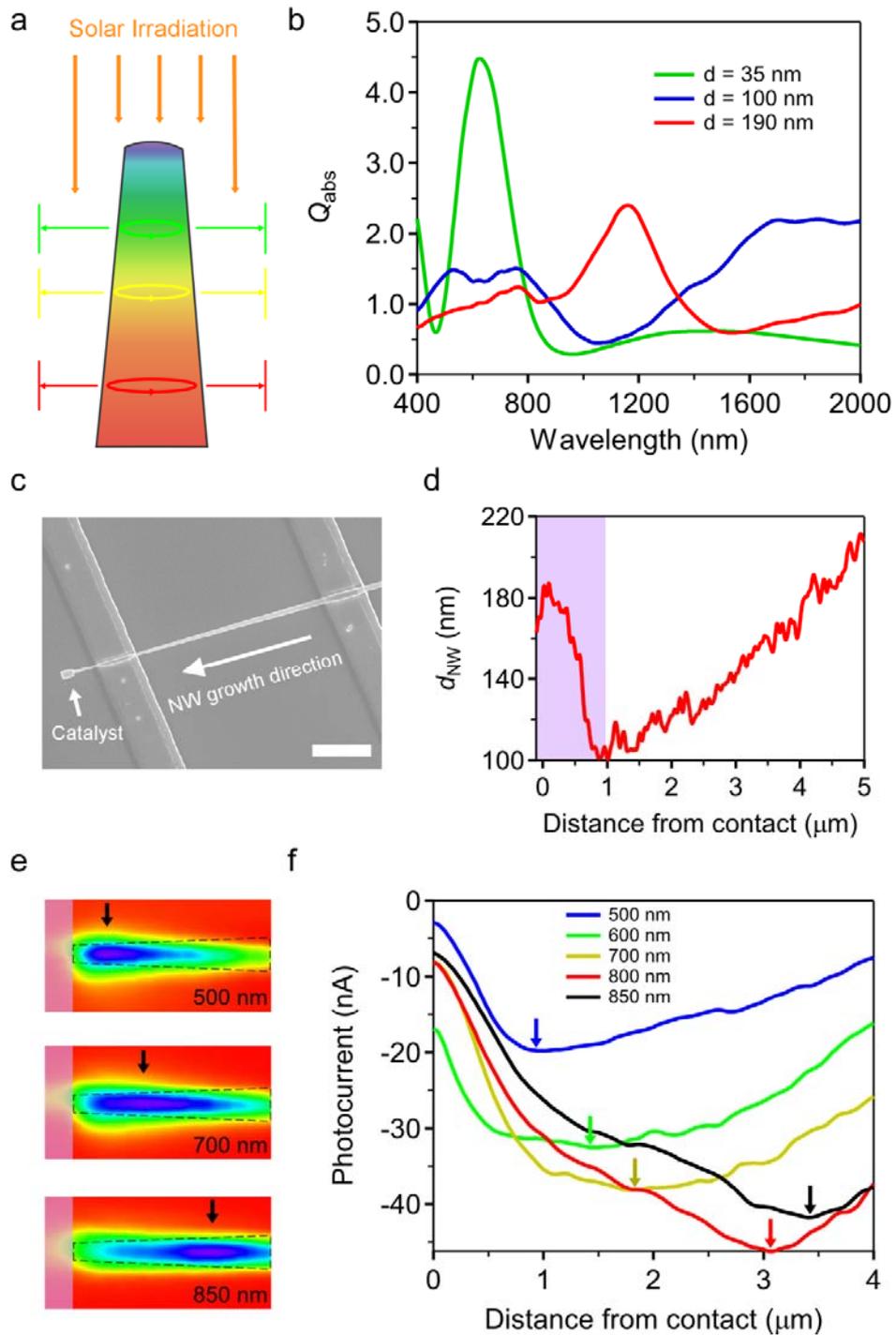

**Figure 2 | Broadband absorption in tapered PbS NWs. a,** Schematic of a tapered NW that absorbs light of different colors at different height, analogous to a tandem cell. Light of longer wavelength is absorbed at larger diameters as expected from the resonance peak shift. **b,** $Q_{abs}$ of

three representative NW diameters of 35, 100, and 190 nm. The resonance enhanced absorption peaks cover the entire solar spectrum including the infrared. **c,** SEM image of a tapered PbS NW device. Scale bar in the inset is 3μm. **d,** Diameter vs. distance from the left contact, as measured by atomic force microscope (AFM). The pink shaded area indicates the contact. **e,** Photocurrent map obtained at $\lambda$ = 500 nm, 700 nm, and 850 nm. The dashed lines in each SPCM image show the tapering direction. Note that the dashed lines are not scaled to the actual NW diameter. **f,** Photocurrent line scan along the NW axis at various incident wavelengths. The photocurrent peak (indicated by arrows) shifts to the right as the wavelength increases.

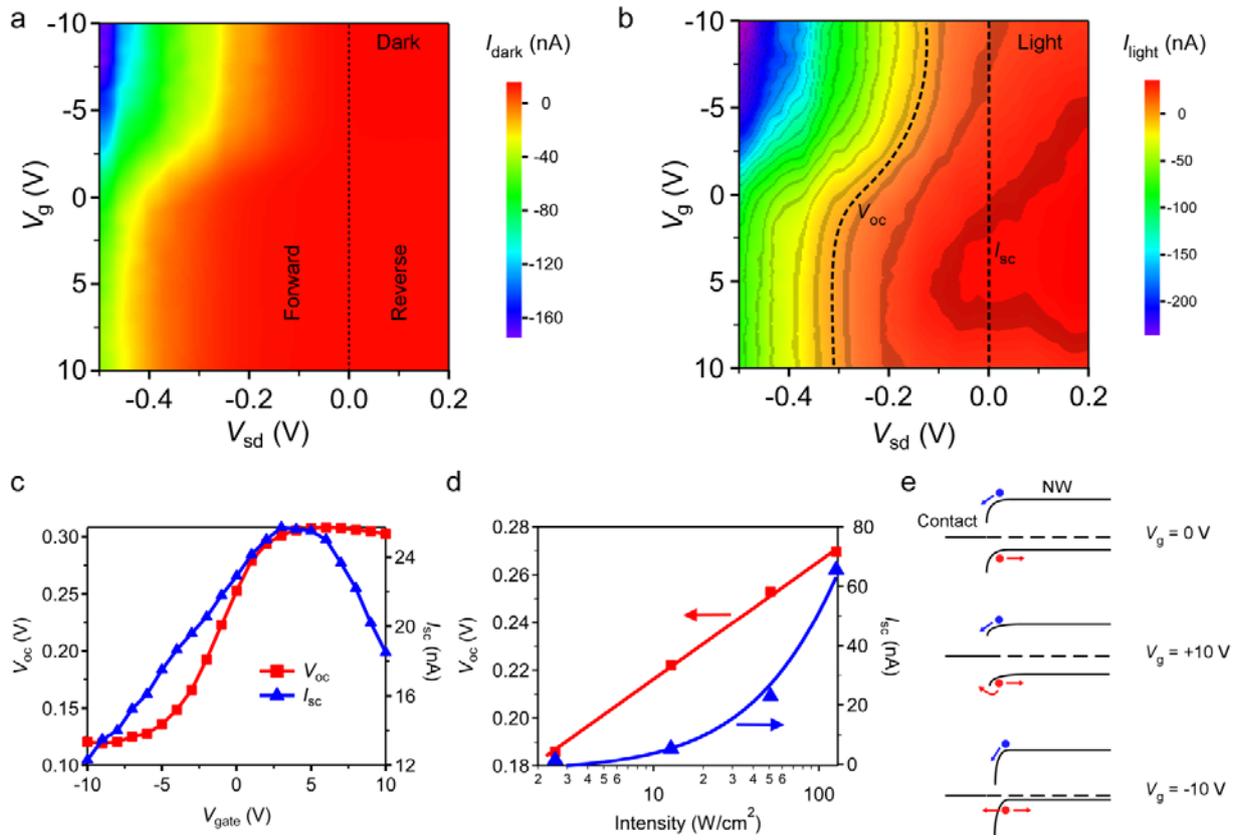

**Figure 3 | Field effects on NW Schottky junctions. a-b,** Current as a function of $V_g$ and $V_{sd}$ in the dark and under the local illumination of 532 nm and 51 W/cm$^2$, respectively. The constant current strips are added in **b** to facilitate map reading. $I$-$V_{sd}$ curves at various $V_g$ are shown in SI. **c,** Gate dependent $V_{oc}$ and $I_{sc}$. **d,** Intensity dependent $V_{oc}$ and $I_{sc}$ at $V_g$ = 0 V. **e,** Schematic band bending diagrams at $V_g$ = 0, +10, and -10 V. Small band bending at $V_g$ = +10 V reduces charge separation efficiency, while large band bending at $V_g$ = -10 V leads to carrier tunneling and an increased dark saturation current (shown in Figure S6).

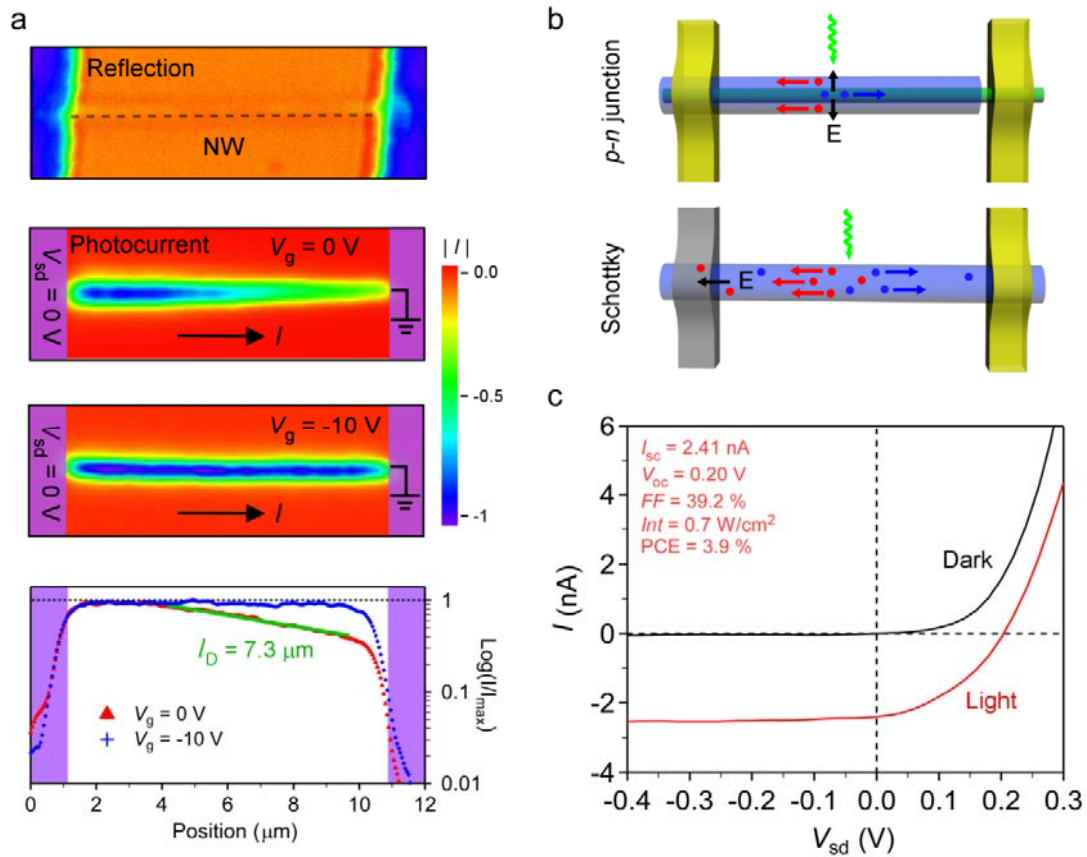

**Figure 4 | Single PbS Schottky solar cell characteristics. a,** Refection, photocurrent images measured at $V_g = 0$V and $V_g = -10$V, and photocurrent line scans along the NW axis. The exponential fitting of the photocurrent line scan yields a long decay length $l_D = 7.3$ μm at $V_g = 0$ V and the photocurrent barely decays over the injection position at $V_g = -10$ V. **b,** Comparison between a core/shell p-n junction NW device and a Schottky NW device where one contact (gray) is Schottky and one contact is Ohmic. Photogenerated carriers in the middle are separated at the radial junction in the core/shell device and diffuse along the NW in the Schottky device. **c,** $I$-$V_{sd}$ characteristics in the dark and under illumination from a halogen bulb in the microscope at 3300K and 0.7 W/cm$^2$.